\documentclass [11pt]{article}

\usepackage[dvips,pagebackref,colorlinks=true,citecolor=blue,linkcolor=blue,pdfpagemode=none,pdfstartview=FitH]{hyperref}

\usepackage{amssymb,amsmath}

\newcommand\R{{\mathbb R}}

\def\epsilon{\varepsilon}
\def\phi{\varphi}
\def\bx{{\bf x}}

\def\bone{{\bf 1}}

\newtheorem{theorem}{Theorem}
\newtheorem{lemma}[theorem]{Lemma}

\newenvironment{proof}{\noindent {\sc Proof:}}{$\Box$ \medskip}

\newcommand\E{\mathop{\mathbb E}\displaylimits}
\renewcommand\P{\mathop{\mathbb P}\displaylimits}

\def\to{\rightarrow}

\title {\bf Is Cheeger-type Approximation Possible for Nonuniform Sparsest Cut?}

\author {{\large\sc Luca Trevisan}\thanks{{\tt trevisan@stanford.edu}.
Computer Science Department, Stanford University. This material is based upon 
work supported by the National Science Foundation under grants No.  CCF 1017403 and CCF-1216642,
and by the US-Israel BSF under grant no. 2010451.}}

\begin {document}

\sloppy

\begin{titlepage}

\maketitle

\begin{abstract} In the {\em nonuniform sparsest cut} problem, given two undirected graphs $G$ and $H$ over the same set of vertices $V$, we want to find a cut $(S,V-S)$ that minimizes the ratio between the fraction of $G$-edges that are cut and the fraction of $H$-edges that are cut. The ratio (which is at most 1 in an optimal solution) is called the {\em sparsity} of the cut.

The special case in which $H$ is a clique over $V$ is called the {\em uniform sparsest} cut problem. In this special case, 
if $G$ is regular, spectral techniques can be used to find a cut of sparsity $O(\sqrt {opt})$, where $opt$ is the sparsity of the optimal cut. We refer to this $O(\sqrt {opt})$-versus-$opt$ approximation as ``Cheeger-type'' approximation. Is such an approximation possible for  the general (nonuniform) problem?

In general, we show that the answer is negative: for every $\epsilon >0$, it is Unique-Games-hard to distinguish 
pairs of graphs for which $opt=1/2$ from pairs of graphs for which $opt=\epsilon$. Furthermore, there is a family of  pairs of graphs such that $opt=1/2$ and the optimum of the Goemans-Linial semidefinite programming relaxation is $o(1)$.

If $H$ is a clique over a subset $V_H$ of the vertices, then Cheeger-type approximation is still impossible using either spectral methods or the Leighton-Rao linear programming relaxation, but we prove that it is achievable by rounding the Gomans-Linial semidefinite programming relaxation.

If $H$ contains the single edge $(s,t)$, then the sparsest cut problem is the minimum s-t-cut problem in $G$. We show that spectral techniques allow a nearly-linear time Cheeger-type approximation. That is, given a network in which the optimal max flow and minimum s-t-cut have cost $\epsilon$ times the sum of the capacities, we can find in nearly linear time a feasible flow and a cut such that the capacity of the cut is at most $O(1/\sqrt {\epsilon})$ times the cost of the flow. (And, in particular, the cut has a capacity that is at most $O(\sqrt \epsilon)$ times the sum of the capacities.)
\end{abstract}\thispagestyle{empty}

\end{titlepage}

\section{Introduction}

Given two undirected weighted graphs $G$, $H$ over the same set of vertices, the {\em sparsity} of a cut $(S,V-S)$ is defined as the ratio of the weight of the edges cut in $G$ over the weight of the edges cut in $H$, both normalized by the total weight:

\[ \sigma(G,H;S) := \frac { \frac {1}{G_{tot}}  \sum_{u,v\in V} G(u,v) \cdot | 1_S(u) - 1_S(v)|} {\frac 1 {H_{tot}} \sum_{u,v \in V} H(u,v) |1_S(u) - 1_S(v)|} \]

where we denote by $G(u,v)$ the weight of the edge $(u,v)$ in $G$, by $H(u,v)$ he weight of the edge $(u,v)$ in $H$, by $G_{tot} = \sum_{u,v} G(u,v)$ the total weight (times 2) of the edges of $G$, by $H_{tot} = \sum_{u,v} H(u,v)$ the total weight (times 2) of the edges of $H$, and by $1_S(\cdot)$ the indicator function of a set $S\subseteq V$. (That  is, $1_S(u)=1$ if $u\in S$ and $1_S(v)=0$ if $v\not\in S$.)

The {\em sparsest cut} problem is, given $G$ and $H$,  to find a cut of minimal sparsity. We denote by $\sigma(G,H)$ the sparsity of a sparsest cut for the graphs $G,H$.

Because of the normalization by $G_{tot}$ and $H_{tot}$, it is easy to see that there always exist a cut of sparsity at most 1, that is, $\sigma(G,H) \leq 1$ for every graphs $G,H$.

An important special case of the sparsest cut problem is the {\em uniform sparsest cut} problem, in which the graph $H$ is a clique with self-loops, that is, $H(u,v) = 1$ for all $u,v \in V$. 

The uniform sparsest cut of a graph $G$ is within a factor of two of the (normalized) {\em Cheeger constant} of $G$, defined as

\[ h(G) := \min_{S \subseteq V} \frac {\sum_{u\in S, v\not\in S} G(u,v) } {\bar d \min \{ |S|, |V-S| \} } \]

where $\bar d :=  G_{tot} /|V|$ is the average degree of $G$.

Another important special case of the sparsest cut problem does not have a standard name, and it corresponds to fixing $H$ to be such that $H(u,v) = d(u) \cdot d(v)$, where $d(x)$ is the (weighted) degree of vertex $x$ in $G$. It can be seen that this parameter is within a factor of two of the {\em conductance} of $G$

\[ {\sc cond} (G) :=  \min_{S \subseteq V} \frac {\sum_{u\in S, v\not\in S} G(u,v) } {\min \{  vol(S),vol(V-S) \} } \]

where $vol(S):= \sum_{v\in S} d(v)$. Note that the conductance is the same as the Cheeger constant for regular graphs, but the two parameters can differ on irregular graphs.

If $G$ is a regular graph, it is known that if the optimum of the uniform sparsest cut problem is $\epsilon$, then it is possible to find in polynomial (and, in fact, nearly linear) time a cut of Cheeger constant at most $\sqrt {2\epsilon}$ and of sparsity at most $\sqrt{8 \epsilon}$. This follows from the discrete {\em Cheeger's inequality}. (This is similar to results proved by Alon and Milman \cite{AM85,A86}; see also the work of Mihail \cite{M89}.) We will refer to an algorithm that finds a cut of sparsity $O(\sqrt {opt})$ where $opt$ is the optimal sparsity as {\em Cheeger-type approximation}.  
Besides the case of uniform sparsest cut in regular graphs, it is also known that, for irregular graphs, Cheeger-type approximation is achievable for the problem of finding  a cut of minimal conductance \cite{Chung96}. (Note that this generalizes the above result.) This gives a Cheeger-type approximation for the version of the non-uniform sparsest cut problem in which $H(u,v) = d(u)d(v)$, where $d(u)$ is the degree of vertex $u$ in $G$.

The algorithms for non-uniform sparsest cut in regular graphs and for conductance in general graphs only require an eigenvector of the second smallest eigenvalue of the normalized Laplacian matrix of the graph, an object that can be computed with good accuracy in nearly-linear time (cf. \cite{ST06:arxiv}), and the cut can be found in nearly-linear time given the eigenvector, or an approximation of it.

We are not aware of a reference establishing a polynomial-time Cheeger-type approximation for the uniform sparsest cut problem and the Cheeger constant problem in irregular graphs. Such an approximation will follow as a special case of our results, although we will not be able to exhibit a nearly-linear-time algorithm.

Another interesting special case arises when $H$ is a graph consisting of a single edge $(s,t)$. Then the non-uniform sparsest cut problem becomes simply the minimum s-t-cut problem in undirected graphs, which is optimally solvable in polynomial time.

In this paper we address the following question: {\em is Cheeger-type approximation possible for the non-uniform sparsest cut problem?}\footnote{Such an approximation would be best possible, because Khot and Vishnoi \cite{KV05} prove than finding a cut of sparsity $opt^{.5 + \epsilon}$ is unique-games hard for every $\epsilon>0$.}

\paragraph{Negative Results.}We provide evidence that the answer is negative, by showing that such an approximation cannot be achieved by the standard semidefinite programming relaxation of sparsest cut, and that it would imply a constant factor approximation for arbitrary instances, a conclusion that, together with the results of \cite{CKKRS05,KV05}, would refute the unique games conjecture.

We then consider a special case that generalizes the known special cases in which Cheeger-type approximation, or even optimal solvability, is known to be doable in polynomial time: we consider graphs $H$ whose non-negative adjacency matrix has rank one, that is, graphs such that there exists an assignment of values $f(v)$ to vertices such that $H(u,v) = f(u) \cdot f(v)$. For example, this includes the case in which $H$ is a clique over a subset of the vertices or the case in which $H$ is a single edge.
We prove that even  the special case in which  $H$ is a clique over a subset of vertices does not admit a Cheeger-type approximation via spectral methods, unlike the case in which $H$ is a clique over all vertices. 

Our reduction from Cheeger-type approximation to constant-factor approximation works by taking an arbitrary instance $(G,H)$ of the non-uniform sparsest cut problem and then constructing another instances $(G,H')$ in which $H'$ is a ``mix'' of $G$ and $H$. The sparsity of a cut in the new instance is essentially the same as in $(G,H)$, but scaled up. The same approach works to convert the Khot-Visnoi integrality gap instances into a family of instances for which the optimum of the sparsest cut is $\Omega(1)$ but the optimum of the Goemans-Linial relaxation is $o(1)$. Our negative result for the spectral approach shows that if $G$ is a ``lollipop graph'' consisting of an expander plus a weighted path leaving from one of the vertices of the expander, and if $H$ is a clique over the vertices of the expanders plus the last vertex of the path in $G$, then the optimum of the spectral relaxation is $o(1)$ even though the optimum of the sparsest cut problem is $\Omega(1)$.

\paragraph{Positive Result for Rank-1 H.} Using the Goemans-Linial semidefinite programming relaxation, however, we show how to achieve Cheeger-type approximation for instances $(G,H)$ in which $H$ is a rank-1 graph. In light of our negative result, the full power of the Goemans-Linial relaxation (that is, both the geometric formulation as a SDP and the presence of triangle inequalities) are needed in order to prove such a result.

Our rounding procedure is rather different from the rounding procedure of Arora, Rao and Vazirani \cite{ARV04} and Arora, Lee and Naor \cite{ALN08}. In particular our (much simpler) rounding procedure does not necessarily produce a Frechet embedding of the negative-type metric coming from the Goemans-Linial relaxation. Rather, we either use a Frechet embedding, in the simple case in which a ``large'' set of points are ``bunched up'' in a small ball, or else we use an embedding from $\ell_2$ to $\ell_1$, the latter case being where we depart from previous work on non-uniform sparsest cut.\footnote{the rounding used by Orecchia and Vishnoi \cite{OV11} for the balanced separator problem is rather similar and so is the proof of Cheeger's inequality for the conductance problem.}

We also give a nearly-linear time Cheeger-type approximation for the minimum s-t-cut in undirected graphs. Our other algorithms are not nearly-linear time because they require a near-optimal solution to the Goemans-Linial relaxation.

\section{Relaxations of the Sparsest Cut Problem}

Given graphs $G,V$, we use the notation $\bar G(u,v) := G(u,v)/G_{tot}$
and $\bar H(u,v) := H(u,v) / H_{tot}$. Note that $\bar G(u,v)$ and $\bar H(u,v)$ are {\em probability distributions} over the set of pairs of vertices in $V$.

Note also that if $H$ is a rank-1 graph, then it is possible to write $\bar H(u,v) = \mu(u) \mu(v)$ where $\mu$ is a probability measure, that is, $H$ being rank-1 is equivalent to the probability distribution $\bar H$ being a {\em product distribution}.

In this paper we will refer to three polynomial-time computable relaxations of the non-uniform sparsest cut problem.

\subsection*{The Leighton-Rao Relaxation}

Leighton and Rao \cite{LR99} studied the following linear programming relaxation

\begin{equation} \label{lr} \begin{array}{lll}
{\rm minimize} & \sum_{u,v} \bar G(u,v) \cdot d(u,v) \\
{\rm subject\ to}\\
& \sum_{u,v} \bar H(u,v) d(u,v) = 1 \\
& d(u,v) \leq d(u,w) + d(w,v) & \forall u,v,w \in V
\end{array}
\end{equation}

\subsection*{The Spectral Relaxation}

\begin{equation} \label{spectral} \min_{x \in \R^V } \frac { \sum_{u,v} \bar G(u,v) \cdot | x_u - x_v |^2} { \sum_{u,v} \bar H(u,v) \cdot | x_u - x_v|^2 }
\end{equation}

We refer to this as a {\em spectral} relaxation because it can be written as
\[ \min_{x\in \R^V} \frac {x^T L(\bar G) x}{x^T L(\bar H) x} \]
where $L(\bar G)$ is the Laplacian of $\bar G$ and $L(\bar H)$ is the Laplacian of $\bar H$.

When $G$ is regular and $H$ is such that $\forall u,v.H(u,v) = 1$  then the above optimization problem is equivalent to

\begin{equation} 
 \min_{x \in \R^V, x \perp \bone } n \cdot \frac { \sum_{u,v} \bar G(u,v) \cdot | x_u - x_v |^2} { 2 x^T x }
\end{equation}

which in turn is the second smallest eigenvalue of the normalized Laplacian of $G$, which can be approximated within an additive error $\delta$ in nearly linear time $O((n+m) \cdot (\log n/\delta)^{O(1)})$ \cite{ST06:arxiv}.

More generally, if $H$ is rank-1 and $\bar H(u,v) = \mu(u)\mu(v)$, then \eqref{spectral} is equivalent to

\[ 
 \min_{x \in \R^V, x \perp \bone } \frac { \sum_{u,v} \bar G(u,v) \cdot | x_u - x_v |^2} { 2n \sum_u \mu(u) x_u^2}
\]

If $H(u,v) = d_u \cdot d_v$, where $d_u$ is the degree of $u$ in $G$, then the above relaxation is again the second smallest eigenvalue of the normalized Laplacian of $G$, which can be approximated in nearly linear time.

Another special case of \eqref{spectral} which can be solved near-optimally in nearly linear time is the case in which $H$ is a graph consisting of a single edge $(s,t)$, as proved in \cite{CKMST11} using the results of \cite{ST06:arxiv}. In such a case, the sparsest cut problem is the {\em undirected s-t-cut problem}. Note that, in this special case, the problem is solvable in polynomial time, and the Leighton-Rao relaxation can be rounded with no loss to provide such an optimal solution. We will return to the minimum s-t-cut in Section \ref{sec:cut}.

As far as we know, it is an open question whether \eqref{spectral} can be solved near-optimally in nearly-linear time for general graphs $G,H$.

\subsection*{The Goemans-Linial Relaxation}

The following semidefinite programming relaxation was proposed by Goemans and Linial:

\begin{equation}\label{gl} \begin{array}{lll}
{\rm minimize} & \sum_{u,v} \bar G(u,v) \cdot || \bx_u - \bx_v ||^2 \\
{\rm subject\ to}\\
& \sum_{u,v} \bar H(u,v) \cdot || \bx_u - \bx_v ||^2 = 1 \\
& ||\bx_u - \bx_w ||^2  \leq ||\bx_u - \bx_v ||^2 + ||\bx_v - \bx_w ||^2 & \forall u,v,w \in V\\
& \bx_u \in \R^d & \forall u \in V\\
& d \geq 1
\end{array} \end{equation}

We recall that if we remove the triangle inequality constraints, then \eqref{gl} becomes equivalent to \eqref{spectral}.
\section{Rounding the Semidefinite Program}

In this section we prove the following result.

\begin{theorem} \label{th:round} There is a polynomial time algorithm that, given in input an arbitrary graph $G$ and a rank-1 graph $H$, finds a cut of sparsity at most $8 \sqrt{opt}$ for $(G,H)$, where $opt$ is the sparsity of an optimal cut.
\end{theorem}

\subsection{Technical Preliminaries}

For a vector $x\in \R^m$ its $\ell_2$ norm is $|| x|| := \sqrt{ \sum_i x_i^2 }$ and its $\ell_1$ norm $||x||_1 := \sum_i |x_i |$. We denote by $\ell_2^m$ the metric space $\R^n$ endowed with the metric $|| x-y||_2$, and we similarly define $\ell_1^m$.

We are interested in embeddings intp $\ell^m_1$, because such embeddings can be translated into feasible solutions for the sparsest cut problem.

\begin{lemma} \label{lm:roundone} Let $G,H$ be an instance of the sparsest cut problem, and let $f: V \to \R^m$ be a mapping of the vertices to real vectors. Then it is possible to find in polynomial time a cut $(S,V-S)$ such that

\[ \frac { \sum_{u,v } G(u,v) |1_S(u) - 1_S(v)| } { \sum_{u,v} H(u,v) |1_S(u) - 1_S(v)|} 
\leq \frac { \sum_{u,v } G(u,v) \cdot ||f(u) - f(v)||_1 } { \sum_{u,v} H(u,v)\cdot || f(u)-f(v)||_1} \] 
\end{lemma}

A useful fact is that, for every dimension $N$,  every $n$ points subsets of $\ell_2^N$ embeds isometrically in $\ell_1^{n\choose 2}$, although the isometry is not known to be efficiently computable. The following weaker result will be sufficient for our purposes.

\begin{lemma} \label{twotoone}For every $n$-point subset $X$ of $\ell_2^N$ there is an embedding $f: X \to \R^m$ such that for every $x,y \in X$

\[ ||f(x) - f(y)||_1 \leq || x - y || \leq 2 \cdot ||f(x)-f(y)||_1 \]
furthermore $m = O(\log n)$ and $f$ can be found with high probability in randomized polynomial time in $n$ and $N$.
\end{lemma}

The factor of 2 in Lemma \ref{twotoone} can be replaced by $1+\epsilon$ for every $\epsilon >0$, although the above result will suffice for our purposes.

Another approach to map metric spaces into $\ell_1^m$ is to use {\em Frechet} embeddings. In a Frechet embedding of a finite metric space $(X,d)$ we select a set $S$, and we define

\[ f_S(x) := \min_{s \in S} d(x,s) \]

Then it is easy to see that  for every $x,y$ we have

\[ | f_S(x) - f_S(y) | \leq d(x,y) \]

\subsection{Proof of Theorem \protect\ref{th:round}}

Leighton and Rao introduced the following idea, which is also used by Arora, Rao and Vazirani: in the {\em uniform} sparsest cut problem, if the metric that we get from our solution is such that $\Omega(n)$ points are concentrated in a ball whose radius is much smaller than the average distance between points, then the points in that ball can be used to derive a Frechet embedding yielding a constant-factor approximation. If there is no small ball containing $\Omega(n)$ points, then $\Omega(n^2)$ of the pairwise distances are at least a constant fraction of the average distance, and this latter fact is useful in developing a rounding procedure.

A version of the above dichotomy holds also when $H$ is a rank-1 graph, instead of a clique. If $H$ is rank-1, then we can write $\bar H(u,v) = \mu(u) \cdot \mu(v)$, where $\mu(\cdot)$ is a probability distribution over $V$, and we have the following fact:

\begin{lemma} \label{dic}
Let $d$ be a semimetric over a set $V$, and $\mu$ be a probability distribution over $V$. Suppose $\E_{u,v \sim \mu} d(u,v) = 1$. Then at least one of the following conditions hold:
\begin{itemize}
\item There is a point $u\in V$ such that the ball $B:= \{ v : d(u,v) \leq 1/4 \}$ is such that
$\mu(B) \geq 1/2$.

\item $\P_{u,v \sim \mu} [d(u,v) \geq 1/4 ] \geq 1/2$
\end{itemize}
\end{lemma}

\begin{proof}
Suppose that the first condition is false, that is, that for every $u$ we have that the ball of radius $1/4$ around $u$ has measure less than 1/2. When we pick a random $u$ and a random $v$ independently, then the probability that $v$ is in the ball of radius $1/4$ around $u$ is less than 1/2, and so there is probability at least $1/2$ and $u$ and $v$ have distance at least $1/4$.
\end{proof}

If the first case of the above lemma holds, then we have a good Frechet embedding.

\begin{lemma}\label{roundf}
Let $\{ \bx_v\}_{v\in V}$ be a feasible solution to the Goemans-Linial relaxation of an instance $(G,H)$ of the non-uniform sparsest cut problem of cost $\epsilon$, suppose that $H$ is a rank-1 graph and let $\bar H(u,v) = \mu(u)\mu(v)$. Suppose that there is $z\in V$ such that $B:= \{  v: || \bx_z - \bx_v||^2 \leq 1/4 \}$ satisfies $\mu(B) \geq 1/2$.

Then, given $G,H, \{ \bx_v\}_{v\in V}, z$ we can find in polynomial time a cut of sparsity
at most $8 \epsilon$.
\end{lemma}

\begin{proof}
See appendix.
\end{proof}

The new part in our analysis is what we do when $d(u,v) := || \bx_u - \bx_v ||^2$ satisfies the second case of Lemma \ref{dic}. In this case we embed the Euclidean distances $|| \bx_u - \bx_v ||$ into $\ell_1$ using Lemma \ref{twotoone}, and we use Cauchy-Schwartz to relate the Euclidean distance-squared to the Euclidean distance.

\begin{lemma}\label{roundcs}
Let $(G,H)$ be an instance of the non-uniform sparsest cut problem, $\{ \bx_v \}_{v\in V}$ be a feasible solution of the Goemans-Linial relaxation of cost $\epsilon$, and suppose that $H$ is a rank-1 graph and that $\mu$ is the distribution such that $\bar H(u,v) = \mu(u) \mu(v)$. Suppose also that 
\[ \P_{u,v\sim \mu} \left[ || \bx_u - \bx_v||^2 \geq \frac 14  \right] \geq \frac 12 \]
Then, given $G,H, \{ \bx_v \}$ we can find in polynomial time a cut of sparsity at most $8\sqrt \epsilon$.
\end{lemma}

\begin{proof} We apply Lemma \ref{twotoone} to find a mapping $f: V \to \R^m$ such that for every $u,v\in V$ we have

\[ || f(u) - f(v)||_1 \leq || \bx_u - \bx_v ||^2 \leq 2 || f(u) - f(v)||_1 \]

Then we have

\[ \E_{(u,v) \sim \bar G} || f(u) - f(v) ||_1 \]
\[ \leq  \E_{(u,v) \sim \bar G}  || \bx_u - \bx_v || \]
\[ \leq  \sqrt{\E_{(u,v) \sim \bar G}  ||\bx_u - \bx_v ||^2 } \]
\[ =  \sqrt{\epsilon  } \]

and

\[ \E_{u,v \sim \mu} || f(u) - f(v) ||_1 \]
\[ \geq \frac 12 \E_{u,v\sim\mu} || \bx_u - \bx_v|| \]

The last assumption in the statement of the lemma is equivalent

\[  \P_{u,v\sim \mu} \left[ || \bx_u - \bx_v|| \geq \frac 12  \right] \geq \frac 12 \]

so that we have

\[  \E_{u,v\sim\mu} || \bx_u - \bx_v|| \geq \frac 14 \]

and so

\[ \frac { \E_{(u,v) \sim \bar G} || f(u) - f(v) ||_1 } { \E_{u,v \sim \mu} || f(u) - f(v) ||_1 } \leq 8 \sqrt \epsilon \]

and, from $f$, we can find a cut of sparsity at most $8 \sqrt \epsilon$.
\end{proof}

\begin{proof}[Of Theorem \ref{th:round}] On input an instance $(G,H)$ of the non-uniform sparsest cut problem in which $H$ is a rank-1 graph. If $\phi(G,H) = \epsilon$ then solving in polynomial time  the Goemans-Linial semidefinite programming relaxation \eqref{gl} will yield a solution $\{ \bx_v \}_{v\in V}$ of cost $\leq \epsilon$. Let $\mu$ be the probability distribution such that $\bar H(u,v) = \mu(u) \cdot \mu(v)$.

By applying Lemma \ref{dic} to the semimetric $d(u,v) := || \bx_u - \bx_v ||^2$ we have that either there is a point $z$ such that $\mu( \{ v: d(u,v) \leq 1/4 \} ) \geq 1/2$ or we have $\P_{u,v \sim \mu} [ d(u,v) \geq 1/4 ] \geq 1/2$. In the former case, we apply Lemma \ref{roundf} to find, in polynomial time, a cut of sparsity $\leq 8 \epsilon \leq 8 \sqrt \epsilon$, and in the latter case we apply Lemma \ref{roundcs} to find, in polynomial time, a cut of sparsity $\leq 8 \sqrt \epsilon$.
\end{proof}

\subsection{An Extension}

If $H$ and $H'$ are two graphs over the same set of vertices, we say that $H'$ is a $(c_1,c_2)$-approximation of $H$ if  for every cut $S$ we have

\[ c_1 \sum_{u,v} H'(u,v) |1_S(u) - 1_S(v) | \leq \sum_{u,v} H(u,v) |1_S(u) - 1_S(v) |
\leq  c_2 \sum_{u,v} H'(u,v) |1_S(u) - 1_S(v) | \]

If $(G,H)$ is an instance of the non-uniform sparsest cut problem whose optimum is $opt$, and
$H$ admits a rank-1 $(c_1,c_2)$-approximation $H'$, then we can find a cut of sparsity at most $8\frac {c_2}{c_1} \sqrt {opt} $ for $(G,H)$ by applying the algorithm of the previous section to $(G,H')$. 

Examples of graphs $H$ that admit good rank-1 approximations are bipartite complete graphs and bounded-degree expander graphs. For such families of graphs $H$, Cheeger-type approximation for instances $(G,H)$ is possible for every $G$.

\section{Max Flow, Min Cut, and the Spectral Relaxation}

If $H$ is a graph consisting of a single edge $(s,t)$, then the relaxation \eqref{spectral} can solved in nearly-linear time. 

\cite{CKMST11} show how to iteratively solve \eqref{spectral} on several instances related to (by different from) the graph $G$ in order to find a nearly-optimal s-t-cut. Here we make the observation, which is probably well known but that we have not seen discussed before, that Cheeger-type approximation for the minimum s-t-cut can be achieved in nearly-linear time.

Suppose that we are given a graph $G$ and a pair $(s,t)$ such that \eqref{spectral} has an optimum solution $x\in \R^V$ of cost $\epsilon$. This means that $|x_s - x_t|^2 = 1$ and

\[ \sum_{u,v} \bar G(u,v) |x_u - x_v|^2 = \epsilon \]

We can assume, by translating the solution if necessary, that $x_s=0$ and $x_t = 1$. Let us construct a random cut by picking a threshold $t$ uniformly at random in the interval $[0,1)$ and defining $S_t := \{ v : x_v \leq t \}$. Then we have that 

\[ \E_t |1_{S_t} (u) - 1_{S_t}(v) | = |x_u - x_v | \]

and so 

\[ \E_t \sum_{u,v} \bar G(u,v) |1_{S_t} (u) - 1_{S_t}(v) |  \]
\[ = \sum_{u,v} \bar G(u,v) |x_u - x_v| \]
\[ = \E_{(u,v) \sim \bar G} |x_u - x_v| \]
\[ \leq \sqrt{ \E_{(u,v) \sim \bar G} |x_u - x_v|^2 } \]
\[ \leq \sqrt{\epsilon } \]

Which means that there is a threshold that yields a cut that is crossed by at most a $\sqrt \epsilon$ fraction of the edges. 

From the standard theory of electrical network, it also follows there is a feasible s-t-flow in the network with capacities $\bar G$ that sends $\epsilon$ units of flow from $s$ to $t$, and that the flow can be found in linear time given an optimal solution $x$. (See \cite{CKMST11} for how to deal with a near-optimal solution $x$.)

\section{Integrality Gap Instances and Unique-Games Hardness}

In this section we prove some negative results about the Cheeger-type approximation
of the non-uniform sparsest cut problem.

Our results are that, in general, Cheeger-type approximation cannot be achieved in polynomial time, unless constant-factor polynomial-time approximation for the non-uniform sparsest cut problem is possible, which would contradict the unique games conjecture. For the Goemans-Linial relaxation \eqref{gl}, we describe integrality gap instances ruling out Cheeger-type approximation via \eqref{gl}.

We then turn to the question of achieving Cheeger-type approximation in the special case in which $H$ is a rank-1 graph. In the even more special case in which $H$ is a clique over a subset of vertices, we show that Cheeger-type approximation is not possible via the spectral relaxation. It was known that Cheeger-type approximation is not possible via the Leighton-Rao relaxation even when $H$ is a clique over all of $V$.

\subsection{General Polynomial Time Algorithms}

\begin{theorem} \label{th:ugc} Suppose that there is a constant $\delta > 0$ and polynomial time algorithm that given an instance $(G,H)$ of the non-uniform sparsest cut problem such that $\sigma(G,H) \leq \delta$ finds a cut of sparsity at most 1/2.

Then there is a polynomial time $1/\delta$-approximate algorithm for the non-uniform sparsest cut problem and the Unique Games Conjecture is false.
\end{theorem}

\begin{proof}
Given a graph $G$ and a cut $S$ of the vertex set $V$ of $G$, we use the notation

\[ G(S) :=  \sum_{u,v} G(u,v) |1_S (u) - 1_S(v)| \]

With the above notation, the sparsity of a cut $S$ for an instance $G,H$ of the non-uniform sparsest cut problem is $\bar G(S)/ \bar H(S)$.

On input an instance $(G,H)$ we guess an $\epsilon$ such that $\epsilon = \phi(G,H)$.
If $\epsilon > \delta$, then we output a cut of sparsity 1, which is a $1/\delta$-approximate solution or better.

Otherwise, we define the instance $(G',H')$ in which $G' = \bar G$ and $H' = \left(1 - \frac \epsilon \delta \right)  \bar G + \frac \epsilon \delta  \bar H$. Then, since $G'$ and $H'$ are already normalized we have that the sparsity of a cut $S$ for this new instance is

\[ \phi(G',H';S) = \frac {G'(S)}{ H'(S) }  \]

Let $S^*$ be an optimal cut for $(G,H)$ of sparsity $\epsilon= \bar G(S^*)/\bar H(S^*)$. Then we have

\[ G'(S^*) = \bar G(S^*) = \epsilon \bar H(S^*) \]
and 
\[ H'(S^*) = \left(1 - \frac \epsilon \delta \right)  \bar G(S^*) + \frac \epsilon \delta  \bar H(S^*)
\geq \frac \epsilon \delta  \bar H(S^*) = \frac 1 \delta G'(S^*) \]
so
\[ \sigma(G',H') = \min_S \frac {G'(S)}{H'(S)} \leq \frac {G'(S^*)} {H'(S^*)} = \delta \]
Applying the polynomial time algorithm in the assumption of the theorem to $(G',H')$ we can find a cut $T$ such that
\[ \frac {G'(T)} {H'(T)} \leq \frac 12 \]
which means that
\[  \left(1 - \frac \epsilon \delta \right)  \bar G(T) + \frac \epsilon \delta  \bar H(T) \geq 2 \bar G(T) \]
that is,
\[ \bar G(T) \leq \frac \epsilon \delta  \bar H(T)  \]
which means that $T$ has sparsity at most $\epsilon/\delta$ for the original instance $(G,H)$, and so is a $1/\delta$-approximate solution.

Khot and Vishnoi \cite {KV05} and Chawla et al. \cite{CKKRS05} prove that  a constant-factor approximation for the non-uniform sparsest cut problem implies that the Unique Games Conjecture is false.
\end{proof}

\subsection{The Semidefinite Programming Relaxation}

The same approach used in the proof of Theorem \ref{th:ugc} shows that given an instance $(G,H)$ of the non-uniform sparsest cut problem such that $\sigma(G,H) = opt$ and the optimum of the Goemans-Linial relaxation is $sdp$,  we can construct an instance $(G',H')$ such that
$\sigma(G',H') \geq 1/2$ and the optimum of the Goemans-Linial approximation is $\leq sdp/2opt$. Khot and Vishnoi  \cite {KV05} describe a family of instances of sparsest cut for which $sdp = o(opt)$, and such a family translates to a family of instances for which $opt \geq 1/2$ and $sdp= o(1)$, proving the Goemans-Linial relaxation cannot lead to a Cheeger-type approximation of the form $opt = \Omega(\sqrt {sdp})$ for general graphs. We proved that such an approximation is possible if $H$ is a rank-1 graph.

\subsection{The Leighton-Rao Relaxation}

If $G$ is a constant-degree expander graph and $H$ is a clique, then it is known that the Leighton-Rao relaxation has an optimum which is $O(1/\log n)$ while the optimum of the sparsest cut problem is $\Omega(1)$. This rules out the use of the Leighton-Rao relaxation to achieve Cheeger-type approximation even if $H$ is a clique (and, in particular, a rank-1 graph).

\subsection{The Spectral Relaxation}

Fix a parameter $k$, and consider the following instance $(G,H)$ over $2k$ vertices $\{ v_1,\ldots,v_{2k} \}$: 

\begin{itemize}
\item $G$ is a constant-degree expander on $\{ v_{k+1}, \ldots,v_{2k} \}$, plus a length-$k$ path  from $v_1$ to $v_{k+1}$;
\item  $H$ is a clique on the $k+1$ vertices $\{ v_1,v_{k+1},\ldots,v_k\}$. 
\end{itemize}

It is easy to see that the sparsest cut is $\Omega(1)$, for example by finding a feasible solution for the dual of the Leighton-Rao relaxation,  but there is a spectral solution of cost $O(1/k)$: set $x_{v_i} = i/k$ for $i\geq k$ and $x_{v_i} = 1$ for $i> k$. 

Then  $G_{tot} = \Theta(k)$, $H_{tot} = \Theta(k^2)$, $\sum_{i,j} G(v_i,v_j) | x(v_i)  - x(v_j)|^2 = O\left(\frac 1 k \right)$, 
and
$\sum_{i,j} H(v_i,v_j) | x(v_i) - x(v_j) |^2 = \Omega(k)$, 
so that the cost of the solution is $O(1/k)$.

This shows that the spectral relaxation cannot achieve a Cheeger-type approximation even if $H$ is a clique on
a subset of nodes (and, in particular, a rank-1 graph).

\newcommand{\etalchar}[1]{$^{#1}$}

\end{document}